# Using GAN to Enhance the Accuracy of Indoor Human Activity Recognition

Parisa Fard Moshiri[1], Hojjat Navidan[1], Reza Shahbazian[1], Seyed Ali Ghorashi[1,2,*] and David Windridge[2]

1: Cognitive Telecommunication Research Group, Department of Telecommunications, Shahid Beheshti University, G.C., Tehran, Iran

2: Department of Computer Science, School of Science and Technology, Middlesex University, London, UK

a_ghorashi@sbu.ac.ir,   *: Corresponding Author

*Abstract*—**Indoor human activity recognition (HAR) explores the correlation between human body movements and the reflected WiFi signals to classify different activities. By analyzing WiFi signal patterns, especially the dynamics of channel state information (CSI), different activities can be distinguished. Gathering CSI data is expensive both from the timing and equipment perspective. In this paper, we use synthetic data to reduce the need for real measured CSI. We present a semi-supervised learning method for CSI-based activity recognition systems in which long short-term memory (LSTM) is employed to learn features and recognize seven different actions. We apply principal component analysis (PCA) on CSI amplitude data, while short-time Fourier transform (STFT) extracts the features in the frequency domain. At first, we train the LSTM network with entirely raw CSI data, which takes much more processing time. To this end, we aim to generate data by using 50% of raw data in conjunction with a generative adversarial network (GAN). Our experimental results confirm that this model can increase classification accuracy by 3.4% and reduce the Log loss by almost 16% in the considered scenario.**

## I. INTRODUCTION

In recent years, activity awareness and recognition have been utilized in various fields of information technology such as Internet of Things (IoT), health diagnosis, fall detection and security of smart houses [1]. In terms of activity recognition, many approaches with motion sensors, smartphones, cameras, or Wi-Fi signals have achieved sustainable results. Wearing motion sensors is inconvenient as it may involve much user effort; battery issues are also a severe weakness for this approach. Gathering video or image signals could be difficult in some particular environments, such as dark rooms. All of these drawbacks encourage researchers to use wireless signals instead of such sensors, involving no additional cost. Whenever a person enters an area with a Wi-Fi transceiver pair within, his/her movements can affect wireless signals; consequently, by analyzing signal characteristics and comparing fluctuations, activities can be recognized. As depicted in Figure 1, it should be noted that different patterns in signals can be observed due to spatial diversity [1].

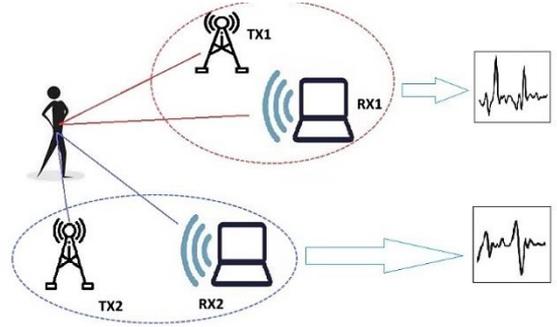

Figure 1.  Two transceiver pairs observe different signal patterns in CSI amplitude for the same activity.

Wireless signals can be described either by Received Signal Strength (RSS) or Channel State Information (CSI). Information about the communication link can be extracted from RSS, while information about the communication channel state is obtained from CSI [2]. The main difference is that RSS is the average value of received signal strength, whereas CSI contains more information about fading channels amplitude and phase [3]. Besides, due to the noisy RSS measurements, it is recommended to use CSI for activity recognition as it contains more stable information [4].

RSS is a successful approach in localization as in whenever a person is located between a Wi-Fi device and an Access Point (AP), the signal is weakened; thus, the value of RSS will be altered. Although RSS is simple to measure, it is incapable of capturing small fluctuations in the signal while a person is moving. The reason behind this incompetence is that RSS is not a stable metric.

As mentioned above, CSI contains both amplitude and phase information, which can be extracted. Each motion will have different effects on CSI; accordingly, each action will have distinct patterns. However, owing to the high noise ratio, CSI measurement cannot be representative enough for different types of human activities [5]. In this circumstance, the best approach is to extract pattern-recognition features. Recently, deep learning algorithms such as Long Short-Term Memory (LSTM), which is able to learn representative features and encode information during learning, has been used for activity recognition with high efficiency [3].

Using LSTMs for activity recognition has two main advantages; First, the LSTM can extract the features automatically without the need for significant pre-processing. Second, LSTM impounds state information of activity temporarily; in other words, it can distinguish between similar compound activities such as "lying down" and "sitting down" because the first activity is composed of latter [3].

The proposed process of this paper can be summarized as follows:

- Initially, we use the benchmark dataset gathered by Siamak Yousefi et al. [7] gathered by three transceiver pairs at 30 subcarriers with PCA for dimensional reduction.

- Afterward, we try to make input features and labels for each activity, such as "fall," "run," "walk," "pick up," "lie down," "sit down" and "stand up." We use the short-time Fourier transform (STFT) to transform the signal into the frequency domain for indicating frequency alteration.

- Since the CSI dataset is large-scale, we try to generate new data set with the same statistical features from data prepared in the first section by using the generative adversarial network (GAN).

- An LSTM is used to classify the activities. A combination of synthetic data with real data improves the accuracy of this model, to approximately 3.4%.

II. RELATED WORKS

Since wireless signals have become pervasive everywhere, many CSI-based human activity recognition systems have been developed. Due to the fading channel and multipath effect, RSS can be extremely noisy. Thus it may not recognize even the simplest activities. In [3] authors propose a deep learning framework for feature selection in the CSI-based human activity recognition (HAR) task. They present an advanced BLSTM network that can learn representative features and encode the temporal information. To put it simply, it can process in both forward and backward directions. Their attention model is a softmax regression layer with normalized weight for each feature and time step. After merging the learned features with the attention matrix in order to flatten the feature matrix to a vector for final classification, a softmax classification layer has been used to recognize different activities.

Fangxin Wang et al. present a WiFi-based spatial diversity-aware device-free activity recognition system, called WiSDAR [8]. They have achieved acceptable results when the human body is appropriately placed between the transceiver pair. When a person moves out of the sweet zone, he enters a dead zone, which can deprecate accuracy. In order to encounter this issue, they identify influential factors. WiSDAR recognizes the dead zones with only one physical transceiver pair. The best approach can be extending multiple antennas for observing activity, even though features from various spatial dimensions are more precious. Therefore, they extract features on time and frequency domain by exerting STFT to generate spectrogram. They categorize all of the generated spectrograms as the initial input in training operation. After preparing data, they use a deep learning model containing Convolutional Neural Network (CNN) and LSTM for integrating features and classifying activities.

Huan Yan et al. propose WiACT, a passive WiFi-based human activity recognition, which exploits the correlation of multiple antennas in order to extract Doppler shift correlation value as input data for an Extreme Learning Machine (ELM) [9]. Since the collected data contains a large amount of noise, the Butterworth low-pass filter is used for noise reduction. They consider that the variance of CSI amplitude during an activity is much higher than the state of no activity. In order to overcome this issue, they design an adaptive activity cutting algorithm (ACAA) by using the difference between the activity and the stationary part invariance feature, adjusting the threshold to achieve the best accuracy. It should be mentioned that the Doppler effect is caused by conversion in the position of the wave source and the observer. For activity classification, a three-layer ELM is used.

Another approach for HAR is to use body sensor data. This type of data can be useful for elderly care or physically impaired people in a smart healthcare environment. In [10], the authors use the Mobile Health dataset (MHEALTHH), collected from ten subjects, by using sign recording of SHIMER2 wearable sensors to achieve HAR. These sensors are placed on the chest, right wrist and left ankle. They have used a hybrid deep learning model consisting of a combination of deep Simple Recurrent Units (SRUs) and Gated Recurrent Units (GRUs). SRUs network processes the sequence of multi-sensors data due to its internal memory capability. Deep GRUs are used to store data and weights in memory to update the values for the future. They also solve the vanishing gradient problem having sufficient information about the quantity of the previously trained data for the future state. Their hybrid model demonstrated excellent performance by achieving accuracy of around 99%.

Jichao Liu et al. [6] propose a Deep Fully Connected network for group activity recognition. Initially, in order to extract moving features for each person, RGB video images are presented to a spatial-temporal model based on LSTM and CNN. The LSTM model can describe the sequence of each action which is a major complement to the CNN observation information. Afterward, they propose the fully connected conditional random field model to analyze the interplay between people in the group, including single-frame images (spatial interaction) and continues multi-frame images (temporal interaction).

Fei Wang et al. propose a deep learning framework for joint activity recognition using CSI fingerprints [11]. Human behavior can affect WiFi signals propagation and demonstrate specific patterns into WiFi signals, called WiFi fingerprints. They have implemented the standard IEEE 802.11n protocol in two USRPs to collect CSI fingerprints. Their collected data consists of 6 activities. Each activity has been repeated for 15 times at 16 different locations. 1-dimensional convolutional network (C1D) has been used to sweep along the time axis of the CSI fingerprints. C1D has two branches for the joint task of localization and activity recognition. It is claimed that their approach is the first manner that views CSI fingerprints as time series, although they obtain an accuracy of 88.13%.

In [12], the authors investigate the performance degradation problem of leave-one-subject-out validation for CSI-based activity recognition using a GAN-based framework named CsiGAN. They conducted experiments on two CSI-based behavior recognition datasets: SignFi data, which contains CSI traces about sign language gestures, and FallDefi data, which includes CSI traces about human activities such as fall, walk, jump, pickup, sit down, and standup. They introduce a new generator to create fake samples, which effectively increases the performance of discriminator. By increasing the number of probability outputs of discriminator from $k+1$ to $2k+1$, where $k$ is the number of classes, they correct the decision boundary for each category. Besides, they propose manifold regularization to stabilize the learning process. However, their method is based on CycleGAN [13]; our work is based on regular GAN and is essentially different. We propose to use GAN for creating synthetic data to achieve the same accuracy with a less number of overall samples, which reduces time and cost, significantly.

## III. SYSTEM MODEL

### A. Channel State Information

Since WiFi mainly exploits Orthogonal Frequency-Division Multiplexing (OFDM), a WiFi can be modeled as a Multiple-Inputs Multiple-Outputs (MIMO). For such a model, CSI is a three-dimensional matrix of complex values representing amplitude and phase of the multi-path channel, where each channel is divided into multiple subcarriers. This whole system is defined as [3]:

$$y_i = H_i x_i + v \text{ for } i = 1, 2, ..., m \quad (1)$$

Where $y_i$ is received signal, $x_i$ is transmitted signal, $H_i$ is the CSI for subcarrier $i$, $n$ is noise term, and $m$ is the total number of subcarriers. In this case, there are one antenna of transmitter and three antennas of receiver; therefore, we can express channel CSI with matrix form as follows:

$$H = \begin{pmatrix} H_{1,1} & H_{1,2} & \cdots & H_{1,30} \\ H_{2,1} & H_{2,2} & \cdots & H_{2,30} \\ H_{3,1} & H_{2,2} & \cdots & H_{3,30} \end{pmatrix} \quad (2)$$

Where $H_{i,j}$ is the CSI value in antenna pair $i$ at $j$-th subcarrier. Since activity recognition is a multiclass single-label classification problem, we make use of cross-entropy or log loss function [14]:

$$L = -\frac{1}{N} \sum_{i=1}^{N} \sum_{j=1}^{C} y_{ij} \log p_{ij} \quad (3)$$

where $N$ is the number of all observations, $C$ is the total number of classes, $y_{ij}$ is a binary indicator, and $p_{ij}$ is predicted probability whether observation $i$ belongs to class $j$.

### B. Pre-processing

Due to the noisy nature of CSI, it is unable to demonstrate features for different activities. Therefore, we should filter out the noise. However, since noises in CSI have high bandwidth, it is impossible to remove them by a typical low pass filter. There are advanced techniques such as principal component analysis (PCA) for this task, which can reduce large dimensions by converting observations of possibly correlated variables into a set of values of linearly uncorrelated variables.

### C. Feature extraction

In order to extract features, the signal is transformed to the frequency domain. This signal transformation can be done by various methods such as short-time Fourier transform (STFT). Since it is applied to each segment, it can detect frequency changes in a signal. For intense activities such as running, we obtain high energy in high frequencies in the spectrogram. These features are employed as the input to the classification algorithm.

### D. Classification

A large number of existing HAR systems use conventional learning techniques such as KNN and HMM in order to classify the extracted feature, although these methods can accost challenges in processing the spectrogram data for a period of time. We use LSTM to classify different activities from the extracted features. Different from CNN that extracts the spatial correlations of inputs, LSTM is mainly used to classify and inherent relationships of time series. It combines the current inputs and the former state stored in memory to exploit time scale relationships. Our model has 200 hidden layers, which keep training until reachable max iterations obtained.

## IV. PROPOSED METHOD

Generative Adversarial Network (GAN), introduced by Goodfellow et al. in 2014 [15], is a deep learning approach towards semi-supervised learning. As depicted in Figure 3, GANs majorly consist of two distinct models; the Generator and the Discriminator. By presenting it with real data, the generator tries to learn the distribution and features of these data and afterward creates synthetic data via input noise with maximum fidelity to the original data. The overall block diagram of a GAN model is depicted in Figure 2. The discriminator is responsible for distinguishing between real data and created data; the result of this process is fed back to the generator to improve its performance. After iterating over multiple times, the generator is finally able to create data such that the discriminator cannot distinguish it from real data. This whole process is summarized in the equation (4) as follows:

$$\min_G \max_D L(D,G) = E_{x \sim p_{data}(x)}[\log D(\mathsf{x})] + \\ E_{z \sim p_z(z)}[\log(1-D(G(\mathsf{z})))] \quad (4)$$

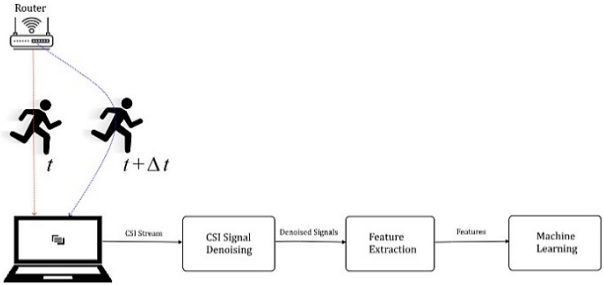

Figure 2. The proposed model consists of: CSI denoising, feature extraction and learning algorithm.

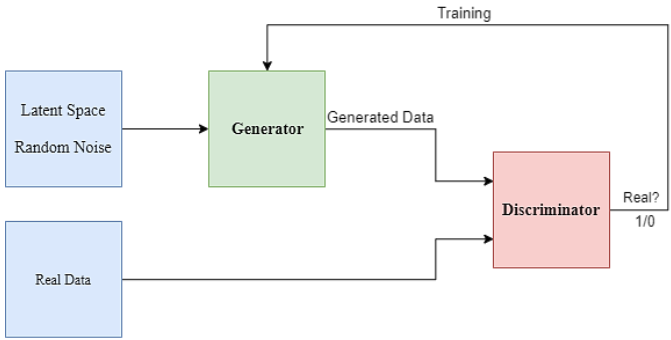

Figure 3. Block diagram of a Generative Adversarial Network. Generated and real data are presented to the discriminator and it's output is used to tune loss functions of both models.

In (4), $D(x)$ is the discriminator's estimation of the probability that $x$ is real, $G(z)$ is the generator's output to input noise $z$, and $D(G(z))$ is the discriminator's estimation of the probability that a generated data is real.

Since the generator is not able directly affect $\log D(x)$ term, minimizing the loss would be equivalent to minimizing the term $\log(1-D(G(z)))$. The cost functions may not converge using gradient descent in a minimax game because when the discriminator becomes optimal, the first term of (4) will vanish. Convergence for a GAN occurs when $D(x) = 0.5$ or in other words, when the discriminator is not able to distinguish between real and generated data which means the generator is finally able to create data with maximum similarity to original data.

## V. EXPERIMENT METHODOLOGY

The dataset used in this paper is provided by Siamak Yousefi et al. [7] and consists of CSI data (both amplitude and phase) for 7 different activities collected at a 1 kHz sample rate. These activities include lying down, falling, walking, running, sitting down and standing up. For classification, we use the same Long Short Term Memory (LSTM) model as proposed in the original paper. Both LSTM and GAN models are implemented on Tensorflow 1.13 and accelerated by Geforce RTX 2060.

We select a large portion of data as our benchmark dataset and evaluate the performance of the classification model on it. Then, 50% of each class of this benchmark dataset is randomly selected and presented to a GAN model to create the missing data. Therefore, each of the 7 classes will contain 50% real data, and 50% synthetic data. The metrics used to evaluate the process are log-likelihood loss and accuracy. The results of this experiment are presented in Table I. As can be seen in Table 1, the accuracy is improved by almost 3%, and the Log Loss has decreased by almost 16% when synthetic data is used beside the real raw gathered CSI. It is evident that this scenario is useful when the data gathering process is so costly, or we are forced to work with a limited amount of data. Otherwise, the accuracy of activity recognition using a complete dataset significantly outperforms the proposed model. In our evaluated model using the complete dataset, the accuracy reaches to 92.8%.

TABLE I. EFFECTS OF ADDING SYNTHETIC DATA ON CLASSIFICATION METRICS.

| Number of Data Samples | | Accuracy | Log Loss |
|---|---|---|---|
| *Real* | *Synthetic* | | |
| 2484 | 0 | 84.3% | 0.48 |
| 2484 | 2489 | 87.2% | 0.40 |
| 4973 | 0 | 92.8% | 0.27 |

## VI. CONCLUSION

CSI based activity recognition is a well-recognized research topic. However, collecting data is time-consuming and costly. Therefore, in this paper, we have demonstrated that the accuracy of CSI-based activity recognition can be improved by appending synthetic data. We proposed to use GAN in order to generate synthetic data, while only a portion of labeled data is used. Our experimental results confirmed that using half of the real data combined with generated synthetic data; the proposed model can improve classification accuracy while reducing the Log Loss.

REFERENCES

[1] Z. Wang, K. Jiang and Y. Hou, "A survey on CSI-based human behavior recognition in through-the-wall scenario," *IEEE Access*, vol. 7, pp. 78772-78793, 2019.

[2] MA. A. Al-qaness, "Device-free human micro-activity recognition method using WiFi signals," *Geo-spatial Information Science*, vol. 22, no. 2, pp. 128-137, 2019.

[3] Z. Chen, L. Zhang, C. Jiang, Z. Cao and W. Cui, "WiFi CSI based passive human activity recognition using attention based BLSTM," *IEEE Transactions on Mobile Computing*, vol. 18, no. 11, pp. 2714-2724, 2018.

[4] W. Zhang, S. Zhou, L. Yang, L. Ou and Z. Xiao, "WiFiMap+: High-level indoor semantic inference with WiFi human activity and environment," *IEEE Transactions on Vehicular Technology*, vol. 68, no. 8, pp. 7890-7903, 2019.

[5] S. Liu, Y. Zhao, F. Xue, B. Chen and X. Chen, "DeepCount: Crowd counting with WiFI via deep learning," arXiv preprint arXiv:1903.05316, 2019.

[6] J. Liu, H. Liu, Y. Chen, Y. Wang and C. Wang, "Wireless sensing for human activity: A survey," *IEEE Communications Surveys & Tutorials, 2019*.

[7] S. Yousefi, H. Narui, S. Dayal, S. Ermon and S. Valaee, "A survey on behavior recognition using WiFi channel state information," *IEEE Communications Magazine*, vol. 55, no. 10, pp. 98-104, 2017.

[8] F. Wang, W. Gong, and J. Liu, "On spatial diversity in WiFi-based human activity recognition: A deep learning-based approach," *IEEE Internet of Things Journal*, vol. 6, no. 2, pp. 2035-2047, 2019.

[9] H. Yan, Y. Zhang, Y. Wang and K. Xu, " WiAct: A passive WiFi-based human activity recognition system," *IEEE Sensors Journal*, 2019.

[10] A. Gumaei, M. M. Hassan, A. Alelaiwi and H. Alsalman, "A hybrid deep learning model for human activity recognition using multimodal body sensing data," *IEEE Access*, vol. 7, pp. 99152-99160, 2019.

[11] F.Wang, et al. "Joint activity recognition and indoor localization with WiFi fingerprints," *IEEE Access*, vol. 7, pp. 80058-80068, 2019.

[12] C. Xiao, D. Han, Y. Ma and Z. Qin, "CsiGAN: Robust Channel State Information-based Activity Recognition with GANs," in *IEEE Internet of Things Journal*.

[13] J.-Y. Zhu, T. Park, P. Isola, and A. A. Efros, "Unpaired image-toimage translation using cycle-consistent adversarial networks," in IEEE International Conference on Computer Vision (ICCV), 2017, pp. 2242–2251.

[14] T. Hastie, R. Tibshirani, J. Frideman and J. Franklin, "The elements of statistical learning: data mining, inference and prediction," *The Mathematical Intelligencer*, vol. 27, no. 2, pp. 83-85, 2005.

[15] I. Goodfellow, et al. "Generative adversarial nets," in *Advances in neural information processing systems*, 2014, pp, 2672-2680.